\documentclass[epj]{svjour}

\usepackage{graphics}
\usepackage{amssymb,amsmath}
\begin{document}
\title{Statistics of power injection in a plate set into chaotic vibration}

\author{Olivier Cadot\inst{1} \and Arezki Boudaoud \inst{2} \and Cyril Touz\'e \inst{1}
}                     
%
%
\institute{ENSTA-UME, Unit\'e de Recherche en M\'ecanique, Chemin
de la Huni\`ere, 91761 Palaiseau, Cedex, France \and Laboratoire
de Physique Statistique, UMR 8550 du CNRS/ENS/Paris 6/Paris 7, 24
rue Lhomond, 75231 Paris Cedex 5, France}
\date{Received: date / Revised version: date}
%
\abstract{A vibrating plate is set into a chaotic state of wave
turbulence by either a periodic or a random local forcing.
Correlations between the forcing and the local velocity response
of the plate at the forcing point are studied. Statistical models
with fairly good agreement with the experiments are proposed for
each forcing. Both distributions of injected power have a logarithmic
cusp for zero power, while the tails are Gaussian for the periodic
driving and exponential for the random one. The distributions of
injected work over long time intervals are
investigated in the framework of the fluctuation theorem, also known
as the Gallavotti-Cohen theorem.  It appears that the conclusions of
the theorem are verified only for the periodic, deterministic forcing.
Using independent estimates of the phase space contraction,
this result is discussed in the light of available theoretical framework.
\PACS{
      {05.40-a}{Fluctuation phenomena, random process, noise and Brownian motion}   \and
      {62.30.+d}{Mechanical and elastic waves; vibrations} \and
      {47.20.Ky}{Non linearity, bifurcation, and symmetry breaking}
     } 
} 
\maketitle
\section{Introduction}
\label{intro}

The statistical distribution of energy and energy fluxes are
central questions concerning out-of-equilibrium dissipative
systems with a large number of degrees of freedom. As energy
fluxes are initiated when forcing the system, it is fundamental to
measure the statistics of the power injected in the system, as in
hydrodynamic fully developed
turbulence~\cite{labbe96,titon03,ciliberto04,titon05}, turbulent
thermal convection~\cite{ciliberto98,aumaitre03}, or turbulent
gravity waves~\cite{falcon08}.

On the theoretical side, the derivation of the fluctuation
theorem~\cite{evans93,gallavotti95,kurchan98} was
a major achievement as it provided, apparently for the first time,
 an exact result to characterize systems far from equilibrium. The
fluctuation theorem (FT), also known as the Gallavotti-Cohen theorem,
essentially requires time-reversibility (in addition to less stringent
conditions not discussed here, see
e.g.~\cite{kurchan07,seifert08,gallavotti08}
for a review). A steady-state
version of the FT can be formulated as follows.
Let $p(t)$ be the instantaneous
injected power, $<p>$ its time-average and $\epsilon_\tau$
its (nondimensional) average over a time interval of
length $\tau$,
\begin{equation*}
    \epsilon _\tau=\frac{1}{\tau}\int_t^{t+\tau}\frac{p(t')}{<p>}dt'.
\end{equation*}
The theorem states the following equivalence for the asymmetry
function
\begin{equation*}
    \rho (\epsilon_\tau)=\frac{1}{\tau} \ln
    {\frac{\pi(+\epsilon_\tau)}{\pi(-\epsilon_\tau)}} \sim \delta \epsilon_\tau
    \quad \textrm{when} \quad \tau \to +\infty,
\end{equation*}
where $\pi(\epsilon_\tau)$ is the probability density function of
$\epsilon _\tau$.

 In the chaotic version of the FT (relying on the chaotic
hypothesis~\cite{evans93,gallavotti95}), $\delta$ is the rate of
contraction in the phase space. In the stochastic
version~\cite{kurchan98,vanzon03b}, $<p>/\delta$ is the
temperature of the stochastic Langevin bath with which the system
is in contact. In this formulation, the FT roughly states that the
probability that the system gives back work
becomes exponentially smaller than the probability
that it takes work
-- if formulated in
terms of entropy, the probability that entropy production is
negative becomes small. Alternatively, this theorem describes how
the system reaches the average behaviour for power injection (or
entropy production).

All experimental verifications of the FT have been performed on
systems with very few (1 to 3) degrees of
freedom~\cite{wang02,schuler05,garnier05,douarche06,jop08}. The
first attempts in turbulent systems~\cite{ciliberto98,ciliberto04}
were hindered by the slow convergence of the statistics and the
lack of events with negative power. As emphasized
in~\cite{aumaitre01}, $\rho(\epsilon_\tau)$ is linear in
$\epsilon_\tau$ for small values of the parameter, so that a
verification of the FT requires reliable measurements up to values
of $\epsilon_\tau$ of order 1. This range was achieved in
gravito-capillary wave turbulence with a random
forcing~\cite{falcon08}, however the asymmetry function
$\rho(\epsilon_\tau)$ turned out to converge to a nonlinear
function when $\tau\to\infty$. This disagreement with the
fluctuation theorem might be ascribed to the lack of
time-reversibility of the equations of hydrodynamics. However, one
might consider instead of the system the ensemble of all its
interacting atoms, for which the equations of motion are
time-reversible. Another plausible explanation stems from the
random forcing. Indeed, the nonlinear $\rho(\epsilon_\tau)$
obtained experimentally~\cite{falcon08} is in agreement with the
prediction of~\cite{farago02} for a particle submitted to viscous
damping and a random Gaussian external force.

A number of questions arise. Is the fluctuation theorem relevant
to turbulent systems where many degrees of freedom are involved~?
What is the role of the nature of the forcing~? To investigate
these questions, we consider a vibrating plate driven with a large
amplitude force, which sets it into a chaotic state of wave
turbulence~\cite{during06,boudaoud08,mordant08}. In this state, a
superimposition of random waves with a broadband spectrum
propagate in the system~\cite{zakharov,newell01}. We drive the
plate with either a deterministic, sinusoidal force or with a
random, Gaussian one. In each case, we investigate the
correlations between the force and the local velocity response of
the plate, the statistics of the power input, and whether the
fluctuation theorem is fulfilled. The article is organized as
follows. In section~\ref{sec:1}, we introduce the experimental
setup. In section~\ref{sec:2}, we give our experimental results
and compare them to theoretical results either from previous work
or derived here. Finally in section~\ref{sec:3}, we discuss our
results which highlight the importance of the nature of the
forcing.

\section{Experiment}
\label{sec:1}

\begin{figure}
\begin{center}
\resizebox{0.8\columnwidth}{!}{%
  \includegraphics{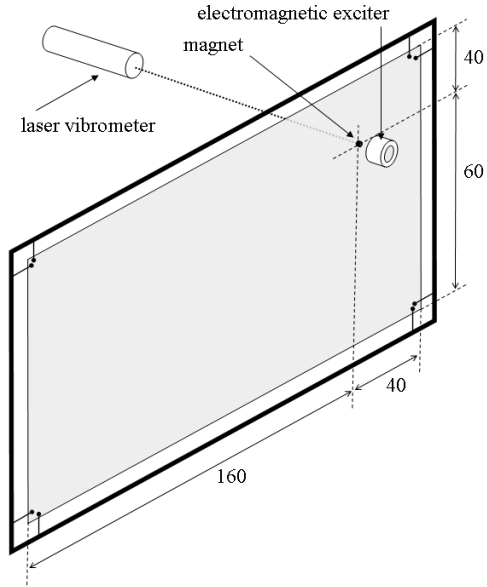}
}
\end{center}
\caption{Sketch of the experimental set-up showing the steel plate
suspended to the frame. Dimensions are in cm. The electromagnetic
exciter (a coil) is placed in front of a magnet. The force acting
on the magnet is controlled by the current in the coil, a laser
vibrometer measures the normal velocity at the other side of the
plate, exactly at the magnet location.} \label{fig:plate}
\end{figure}

\begin{figure}
\begin{center}
\resizebox{0.75\columnwidth}{!}{%
  \includegraphics{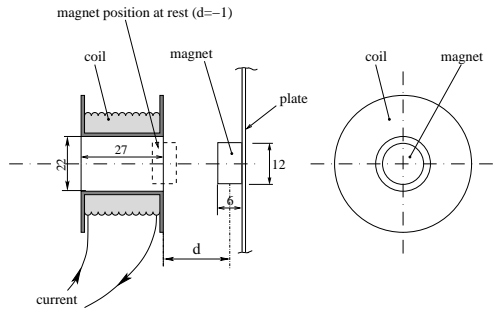}
}
\end{center}
\caption{Electromagnetic exciter. The magnet is radially centred
with the axis of the coil cavity. The dashed rectangle represents
the magnet position when the plate is at rest ($d=-1$). Dimensions
are in mm.} \label{fig:exciter}
\end{figure}

 The experimental set-up is the same as in
\cite{boudaoud08}, except that the excitation device is now
replaced by a magnet-coil system. The vibrating plate is a steel
plate suspended by each of its corners to a rigid frame (see
Figure~\ref{fig:plate}). The plate was chosen for its very high
modal density, obtained by large dimensions $2\;\textrm{m}\times
1\;\textrm{m}$ for a thickness of $h=0.5$ mm. Material properties
were estimated as: Young's modulus $E=200$ GPa, Poisson's ratio
$\nu=0.3$ and mass per unit volume $\rho=7800\, \textrm{kg/m}^3$.
The forcing device consists of a coil and a permanent magnet
simply magnetized on the steel plate (see Figure~\ref{fig:plate}
and \ref{fig:exciter}). When the plate is at rest, the magnet
position is $d=-1mm$ as depicted by the dashed rectangle in
Figure~\ref{fig:exciter}. It has been shown in \cite{thomas03}
that in this configuration, the force $F$ acting on the magnet is
proportional to the current $I(t)$ circulating in the coil
$F(t)=KI(t)$. The current is measured by inserting an ohmic
resistance of $0.12 \Omega$ in series with the coil. For the
calibration, we used the previous measurements in
\cite{boudaoud08}, where the excitation was performed with a
shaker mounted with a force sensor. The proportionality constant
$K$ was estimated by comparing the measured fluctuating current
(when the plate is forced by the electromagnetic exciter with
sinusoidal current) to the measured fluctuating force (when the
plate is forced by the shaker with a sinusoidal tension) for the
same fluctuating normal velocity of the plate at the application
point of the forcing. The proportionality constant found is
$K=0.456$N/A. The normal velocity $v$ at the application point of
the forcing is measured with a laser vibrometer from Polytec
(model OFV 056), as depicted in Figure~\ref{fig:plate}. The normal
velocity $v$ and the coil intensity $I$ are simultaneously
acquired at the sampling frequency of 2000 Hz.

We present two experiments with fundamentally different forcing. A
signal generated by a PC is amplified with a QSC audio RMX 2450
professional power amplifier which supply the coil. For the first
experiment, referred to as the periodic forcing, the signal is
sinusoidal with a frequency of $75Hz$. For the second experiment,
referred to as the random forcing, the signal is a white noise in
the range $0.001Hz$ to $75Hz$, but the resulting force on the
magnet is modified by the frequency response of the amplifier. The
auto power spectrum $P_F(f)$ of the resulting force $F$ is shown
in Figure~\ref{fig:Spectre-F,v}(a) for the periodic forcing and
\ref{fig:Spectre-F,v}(b) for the random forcing. For each case,
the velocity response is chaotic. The power spectrum $P_v(f)$ in
Figure~\ref{fig:Spectre-F,v} covers a wide range of frequencies.
The part for the high frequencies ($f>75Hz$) is insensitive to the
nature of the forcing, and corresponds to the domain of wave
turbulence investigated in~\cite{boudaoud08,mordant08}.

In order to obtain enough statistics for data processing, the
acquisition duration was 15 hours for the periodic forcing and 9
hours for the random forcing. For each case, the amplitude of the
force was chosen to be sufficiently high to reach the wave
turbulence state but low enough to avoid excessive heat that could
damage the ohmic resistance of $0.12\Omega$ during the long
acquisition. The circuitry (coil and resistance) was also
ventilated with a fan for cooling to insure a steady state.


\begin{figure*}
\begin{center}
\resizebox{2\columnwidth}{!}{%
  \includegraphics{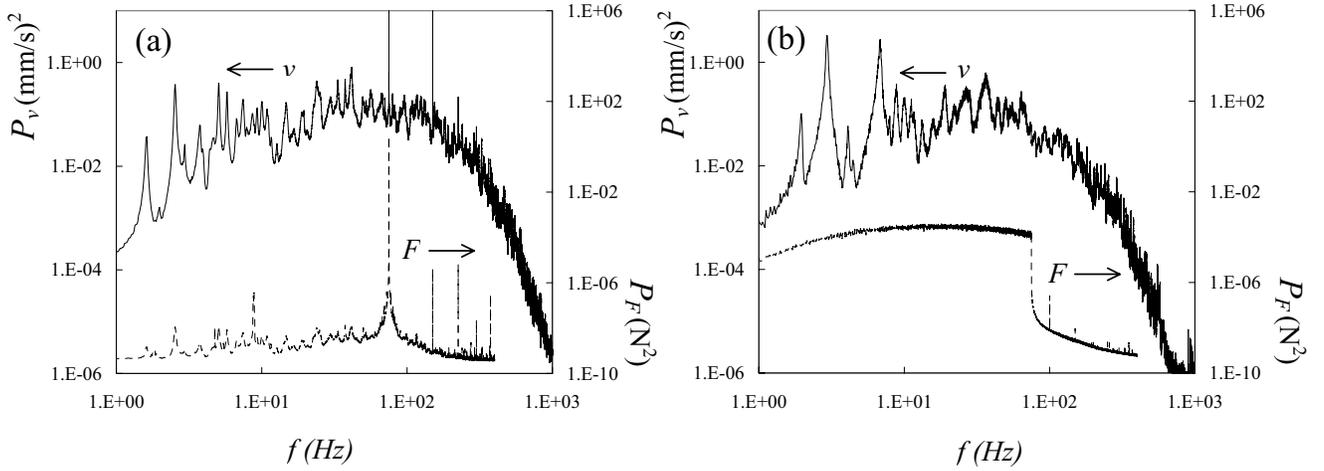}
}
\end{center}
\caption{Power spectrum of the force $F$ and the normal
velocity $v$ for the periodic forcing (a) and the random forcing
(b).} \label{fig:Spectre-F,v}
\end{figure*}

\section{Results}
\label{sec:2}

\subsection{Statistics of the forcing}

\begin{figure}[h]
\begin{center}
\resizebox{0.9\columnwidth}{!}{%
  \includegraphics{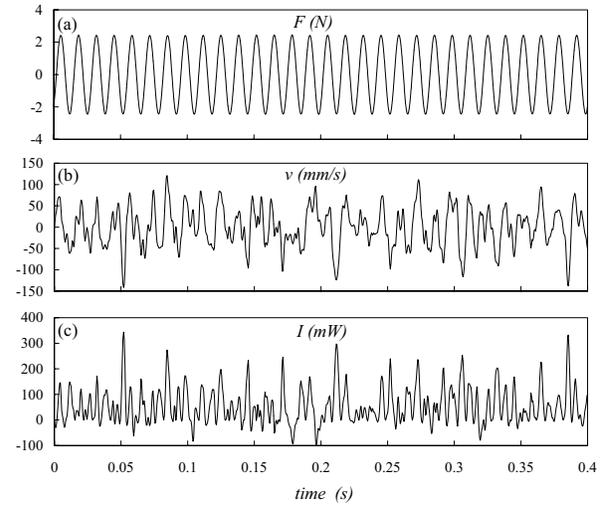}
}
\end{center}
\caption{Excitation with a periodic force. Time series of the
local force (a), normal velocity at the forcing point (b) and
injected power in the plate (c).} \label{fig:timeseries-sinus}
\end{figure}

\begin{figure}
\begin{center}
\resizebox{0.9\columnwidth}{!}{%
  \includegraphics{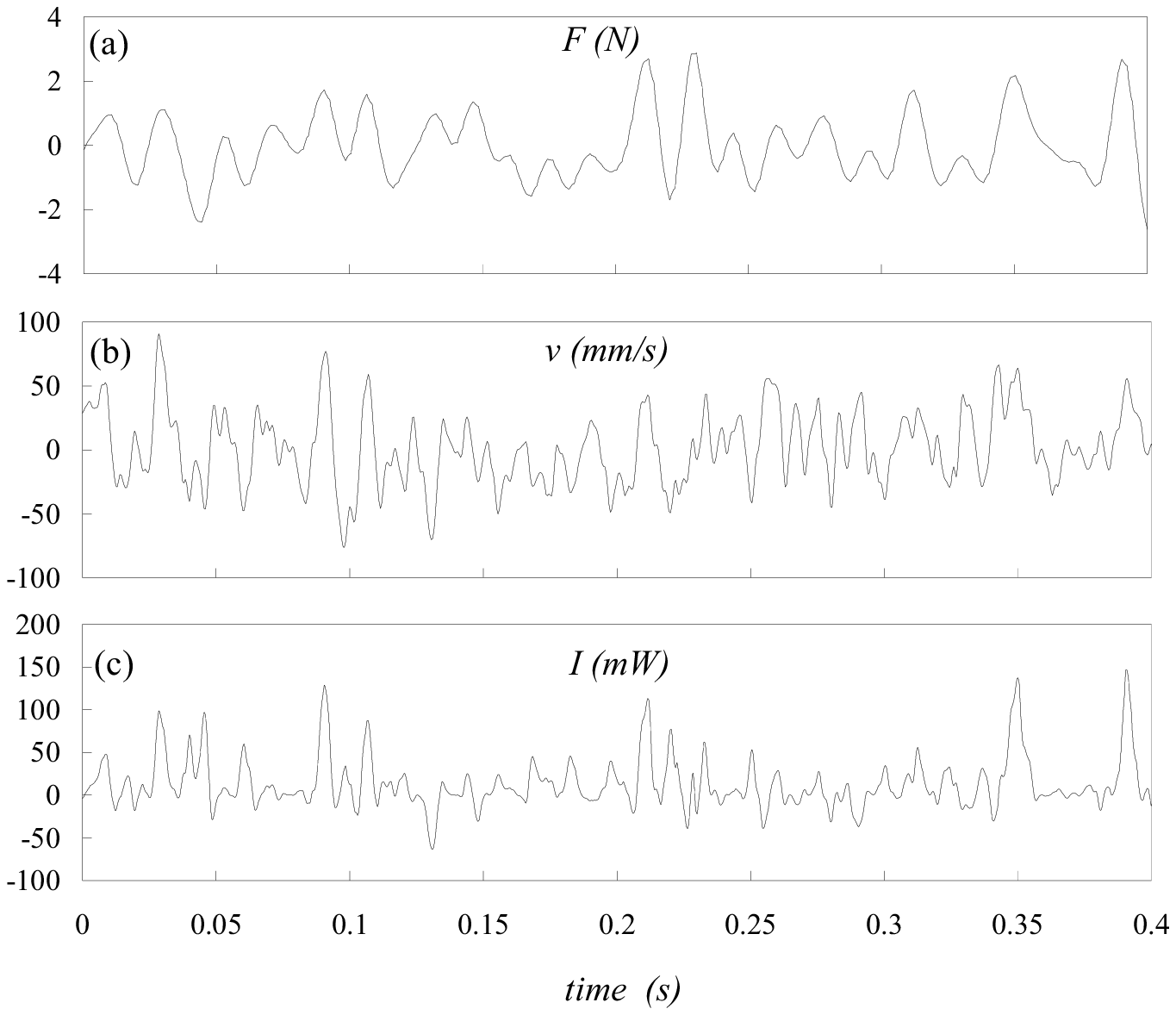}
}
\end{center}
\caption{Excitation with a random force. Time series of the local
force (a), normal velocity at the forcing point (b) and
injected power in the plate (c).} \label{fig:timeseries-random}
\end{figure}

Time series of the force $F(t)$, velocity $v(t)$ and injected
power $p(t)=F(t)\cdot v(t)$ are displayed
Figure~\ref{fig:timeseries-sinus} for the periodic forcing and in
Figure~\ref{fig:timeseries-random} for the random forcing. The
force is computed in Newton from the current circulating in the
coil as explained in section~\ref{sec:1}. For each case the normal
velocity of the plate (Figure~\ref{fig:timeseries-sinus}b and
Figure~\ref{fig:timeseries-random}b) presents a chaotic behavior
corresponding to the wave turbulence regime as described in
\cite{boudaoud08,mordant08}.

In the following, $<x>$ denotes the time average of the variable
$x$ and $\sigma_x = \sqrt{<(x-<x>)^2>}$ its standard deviation.
The statistics are performed over the whole duration of the
experiments. Table~\ref{tab:mean-standard-forcing} summarizes the
main statistical magnitudes of the force, velocity and input power
of both experiments.
\begin{table}
\caption{Average and standard deviations of power $p$ (mW), force
$F$ (N) and normal velocity $v$ (mm/s) related to the periodic and
random forcing. The correlation coefficient of $F$ and $v$ is
$r=\frac{<Fv>}{\sigma_F \sigma_v}=\frac{<p>}{\sigma_F \sigma_v}$.}
\label{tab:mean-standard-forcing}
\begin{tabular}{llllll}
\hline\noalign{\smallskip}
Forcing & $<p>$ & $\sigma_p$ & $\sigma_F$ & $\sigma_v$ & $r$  \\
\noalign{\smallskip}\hline\noalign{\smallskip}
periodic & 54.34 & 145.2 & 1.72 & 46.52 & 0.680\\
random & 18.9 & 107.2 & 1.12 & 31.73 & 0.531\\
\noalign{\smallskip}\hline
\end{tabular}
\end{table}
Whatever the forcing, the force in
Figure~\ref{fig:timeseries-sinus}(a) or
\ref{fig:timeseries-random}(a), and the normal velocity in
Figure~\ref{fig:timeseries-sinus}(b) or
\ref{fig:timeseries-random}(b), fluctuate around a zero temporal
mean value. Although the forcing is stronger for the periodic case
than for the random case, the ratio of the response to the
excitation $\sigma_v/\sigma_F$ are equivalent; $0.27$~(m/s)/N for
periodic forcing and $0.28$~(m/s)/N for the random forcing. The
variables are strongly correlated, the correlation coefficient $r$
between $F$ and $v$ (see Table~\ref{tab:mean-standard-forcing}),
$r=\frac{<Fv>}{\sigma_F \sigma_v}=\frac{<p>}{\sigma_F \sigma_v}$
is larger for the periodic case than for the random case. The
injected power in Figures~\ref{fig:timeseries-sinus}(c) and
\ref{fig:timeseries-random}(c) present intermittent fluctuations
around non-zero and positive mean values corresponding to the
dissipation of the system. We find for mean injected power (i.e.
mean dissipation), $<p> =54.4$ mW for the periodic forcing and
$<p>=18.9$ mW for the random forcing. For both cases, the
fluctuation rate computed from
table~\ref{tab:mean-standard-forcing} is very large;
$\sigma_p/<p>=267\%$ for periodic forcing and $567\%$ for random
forcing. Because the dissipation is small compared to the power
fluctuations, negative events of injected power are often
observable,  more frequently for the random forcing than for the
periodic forcing.

\begin{figure*}
\begin{center}
\resizebox{2\columnwidth}{!}{%
  \includegraphics{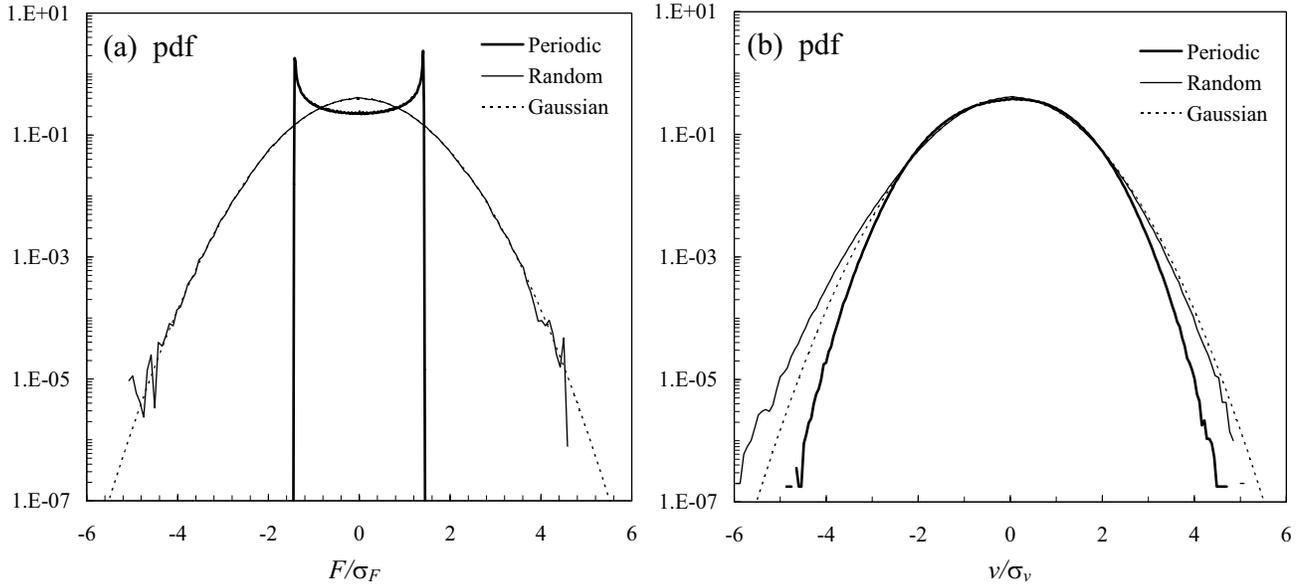}
}
\end{center}
\caption{Probability density functions of the reduced force
$F/\sigma_F$ (a) and reduced normal velocity $v/\sigma_v$ (b) for
periodic forcing (thick line) and random forcing (thin line). The
dashed line is a Gaussian statistic.} \label{fig:Pdf-F,v}
\end{figure*}
\begin{figure*}
\begin{center}
\resizebox{2\columnwidth}{!}{%
  \includegraphics{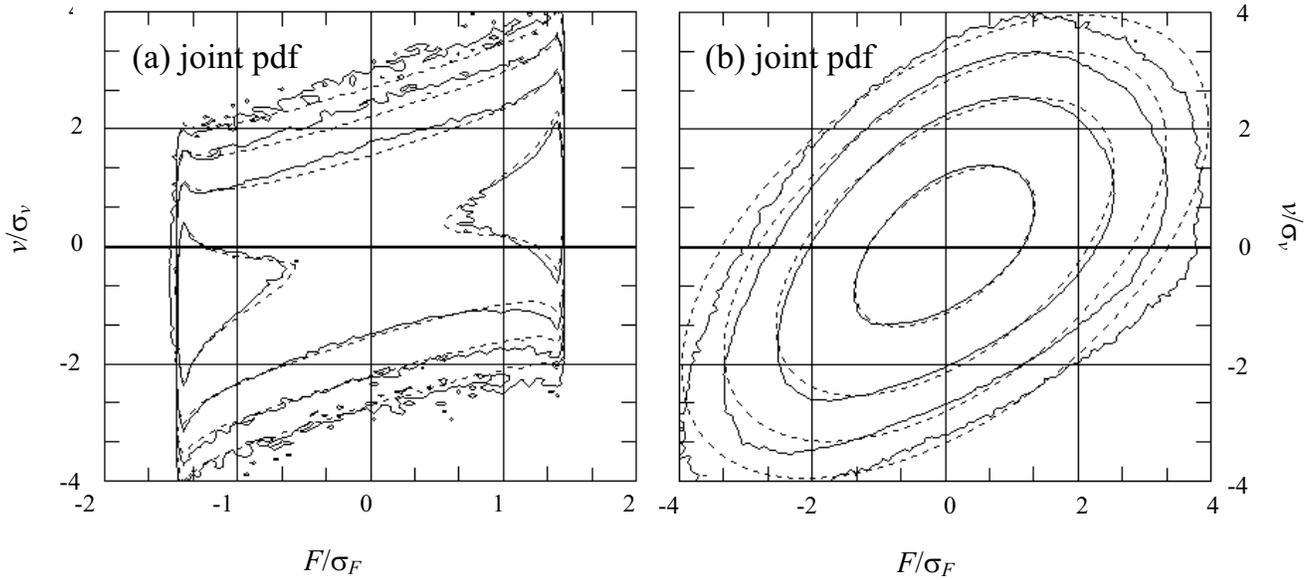}
}
\end{center}
\caption{Joint pdf of the force $F$ and the normal velocity $v$
for the periodic forcing (a) and random forcing (b). The
continuous lines correspond to the experiments. The dashed lines
in (a) correspond to the periodic model (see text) and in (b) to
the bivariate Gaussian distribution (see text) for same
correlation coefficient $r$ and standard deviations $\sigma_F$ and
$\sigma_v$. The isoprobability contours are logarithmically
spaced, and are separated by factors of 10.}
\label{fig:JointPdf-F,v}
\end{figure*}

\begin{figure*}
\begin{center}
\resizebox{2\columnwidth}{!}{%
  \includegraphics{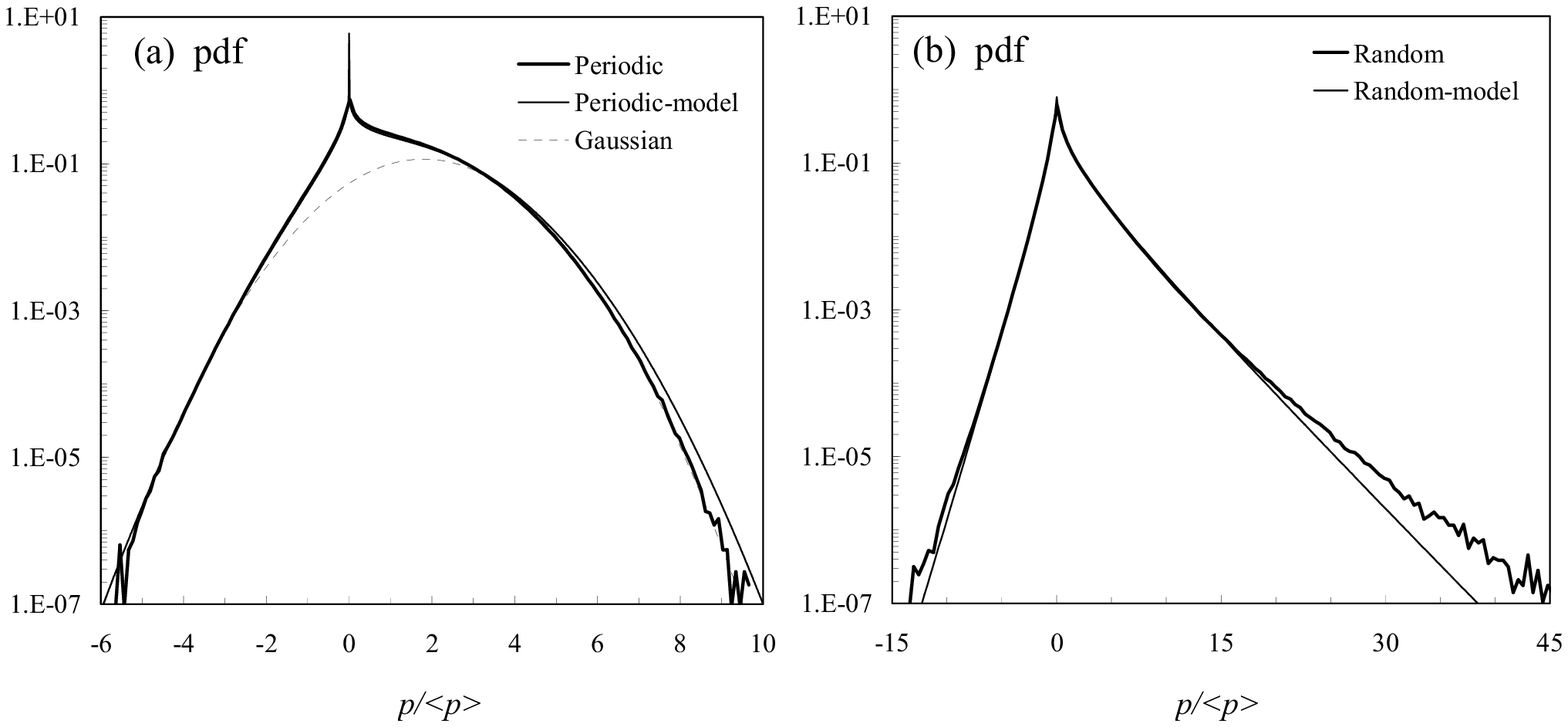}
}
\end{center}
\caption{Probability density functions of the reduced injected
power $p/<p>$ for the periodic forcing (a) and random forcing (b).
Experimental data (thick line) and model (thin line, see text) are
almost undistinguishable (no fitting parameter); the dashed line
is a Gaussian.} \label{fig:Pdf-p}
\end{figure*}%

The pdf of the force is plotted in Figure~\ref{fig:Pdf-F,v}(a) for
both types of forcing. As imposed by the exciter, we recover the
pdf of a sine function for the periodic case, and a Gaussian
distribution for the random forcing. The response of the system is
shown in Figure~\ref{fig:Pdf-F,v}(b). We can see that the pdfs of
the normal velocity are rather close to Gaussian. The deviation
from the Gaussian depends on the type of forcing: the pdf is
symmetric and sub-Gaussian for the periodic forcing and skewed
(with an excess of negative events) for the random forcing. While
the imperfections of the plate \cite{boudaoud08} can explain the
asymmetry in the velocity fluctuations response, it appears that
the random forcing is more sensitive to these imperfections than
the periodic forcing.

The joint pdfs of the force $F$ and the normal velocity $v$ are
shown in Figure~\ref{fig:JointPdf-F,v}. They are drastically
different, but both exhibit a strong positive correlation
between $F$ and $v$. They are compared to statistical models
(dashed lines) that will be presented and discussed in next sub-section.
Likewise, the difference in injected power statistics
is striking, as displayed in  Figure~\ref{fig:Pdf-p}. For the periodic
forcing in Figure~\ref{fig:Pdf-p}(a), the tails for both positive
and negative events match asymptotically the Gaussian statistics
while for the random forcing in Figure~\ref{fig:Pdf-p}(b), the tails
are exponential, with a strong asymmetry of
positive skewness. These tails show that large
fluctuations are regular for the periodic forcing and intermittent
for the random forcing. This fundamental difference between the two
types of forcing can be recovered by the statistical models discussed
hereafter.

\subsection{Statistical models for the injected power}
\subsubsection{Periodic model}  \label{sec:Periodic model}
When the plate is periodically forced with
\begin{equation}\label{F-sinus}
F = \sigma_F \sqrt 2 \sin(2\pi f t),
\end{equation}
one can view the normal velocity response $v$ as the sum of a
sinusoidal time-dependent variable related to the forcing and a
Gaussian variable $u$ related to wave turbulence. We then simply
write the sinusoidal variable as proportional to the sinusoidal
forcing
\begin{equation}
v=u + \frac{r \sigma_v}{\sigma_F}F, \label{v-sinus}
\end{equation}
where the prefactor of $F$ has been adjusted so that
${<p>}=\-<Fv>=r\sigma_v \sigma_F$, assuming that $u$ and $F$ are
independent. Taking the average of the square of (\ref{v-sinus}),
we obtain a relation between the standard deviations of $u$ and
$v$,
\begin{equation}
\sigma_v^2=\frac{\sigma_u^2}{1-r^2}. \label{sigma v-sinus}
\end{equation}
The injected power can be written as
\begin{equation}
p=vF=uF+ \frac{r \sigma_v}{\sigma_F}F^2.
\end{equation}
Then, introducing
\begin{equation}
q=\frac{r}{\sqrt{1-r^2}},\label{q}
\end{equation}
the pdf of $p$ can be computed as
\begin{eqnarray}
\lefteqn{f_p(\tilde{p}=p/<p>) =} \nonumber\\ &   &
\int f_F(\tilde{F})f_u(\tilde{u}) \  \delta
\left(\tilde{p}-\frac{1}{q}\tilde{F}\tilde{u}-\tilde{F}^2\right) \
\mathrm{d}\tilde{F} \ \mathrm{d}\tilde{u} \label{pdf-p-sinus-1}
\end{eqnarray}
where
\begin{equation}
f_v(\tilde{u}=u/\sigma_u)=\frac{1}{\sqrt{2\pi}}\exp\left(-\frac{\tilde{u}^2}{2}\right)
\end{equation}
is the pdf of the Gaussian variable and
\begin{equation}
f_F(\tilde{F}=F/\sigma_F)=\frac{1}{\pi\sqrt{2-\tilde{F}^2}}
\end{equation}
the pdf of the sinusoidal force. Therefore,
(\ref{pdf-p-sinus-1}) becomes
\begin{eqnarray}
\lefteqn{f_p(\tilde{p}=p/<p>) =} \nonumber\\ &
\displaystyle{\sqrt{\frac{1}{\textbf{2}\pi^3}} \int_0^{\sqrt{2}}
\frac{q}{\tilde{F}\sqrt{2-\tilde{F}^2}}
\exp\left(-\frac{q^2}{2}\left(\frac{\tilde{p}}{\tilde{F}}-
\tilde{F}\right)^2\right)\ \mathrm{d}\tilde{F}.}
\end{eqnarray}
Apparently, it cannot be reduced to elementary functions. However,
this pdf has a Gaussian tail,
\begin{equation}
\log f_p(\tilde{p})\sim-\frac{1}{4}q^2\tilde{p}^2
\quad \textrm{when}\quad \tilde{p}\to\infty,
\end{equation}
 while at 0, this pdf has a logarithmic cusp,
 \begin{equation}
 f_p(\tilde{p}) \sim \frac{q}{\pi^{3/2}}\log\left(\frac{1}{\tilde{p}}\right)
 \quad \textrm{when}\quad \tilde{p}\to 0 .
 \end{equation}
Therefore, it is most likely to inject zero power in the system.
For comparison with the experimental data, we only need the
experimental values given in Table~\ref{tab:mean-standard-forcing}
(implying $q=0.926$) as inputs of the model; there is no curve
fitting. The joint pdf of $F$ and $v$ of this statistical model is
the dashed line in Figure~\ref{fig:JointPdf-F,v}(a) and the pdf of
injected power is the thin line called ``Periodic model" in
Figure~\ref{fig:Pdf-p}(a). The periodic model agrees very
satisfactorily with the experimental data.

\subsubsection{Random model} \label{sec:Random model}
For the random forcing, a similar shape to
Figure~\ref{fig:Pdf-p}(b) has already been obtained and modelled
in three-dimensional hydrodynamic developed
turbulence~\cite{pumir96} and in two hydrodynamical systems
driven by a random force: gravito-capillary wave
turbulence~\cite{falcon08} and two-dimensional hydrodynamic
turbulence~\cite{bandi08}. We compare in
Figure~\ref{fig:JointPdf-F,v}(b), the joint pdf of the force and
the velocity to the bivariate Gaussian distribution, having the
same correlation coefficient $r$ and standard deviation $\sigma_v$
and $\sigma_F$ as  in the experiment (see
table~\ref{tab:mean-standard-forcing}). The agreement is fairly
good except for simultaneous large events of same sign of $F$ and
$v$. These events corresponding to the largest fluctuations of
power injection, are underestimated by the bivariate Gaussian
distribution for large events of negative velocity and
overestimated for large events of positive velocity. This shift is
related to the negative skewness of the experimental velocity
distribution as displayed in Figure~\ref{fig:Pdf-F,v}(b). An
analytical expression of the pdf of the injected power can be
computed \cite{falcon08} from the Gaussian bivariate distribution
\begin{eqnarray}
\lefteqn{f_p(\tilde{p}=p/\sigma_p)=\frac{1}{\pi\sigma_v\sigma_F\sqrt{1-r^2}}}
\nonumber\\ &   &
\exp\left(\frac{r\tilde{p}}{(1-r^2)\sigma_u\sigma_F}\right)K_0\left(\frac{\tilde{p}}{(1-r^2)\sigma_v\sigma_F}\right).
\end{eqnarray}
Again, this distribution has a logarithmic cusp at zero power;
in contrast, its tails are exponential.
The pdf of the injected power is the thin line plotted in
Figure~\ref{fig:Pdf-p}(b). Again, there is no curve fitting, and
only the data given in Table~\ref{tab:mean-standard-forcing} are
needed. The model agrees very satisfactorily with the experiment, except
again for the large power positive deviation for the reason
mentioned above.

\subsection{Fluctuation theorem}

\begin{figure*}
\begin{center}
\resizebox{2\columnwidth}{!}{%
  \includegraphics{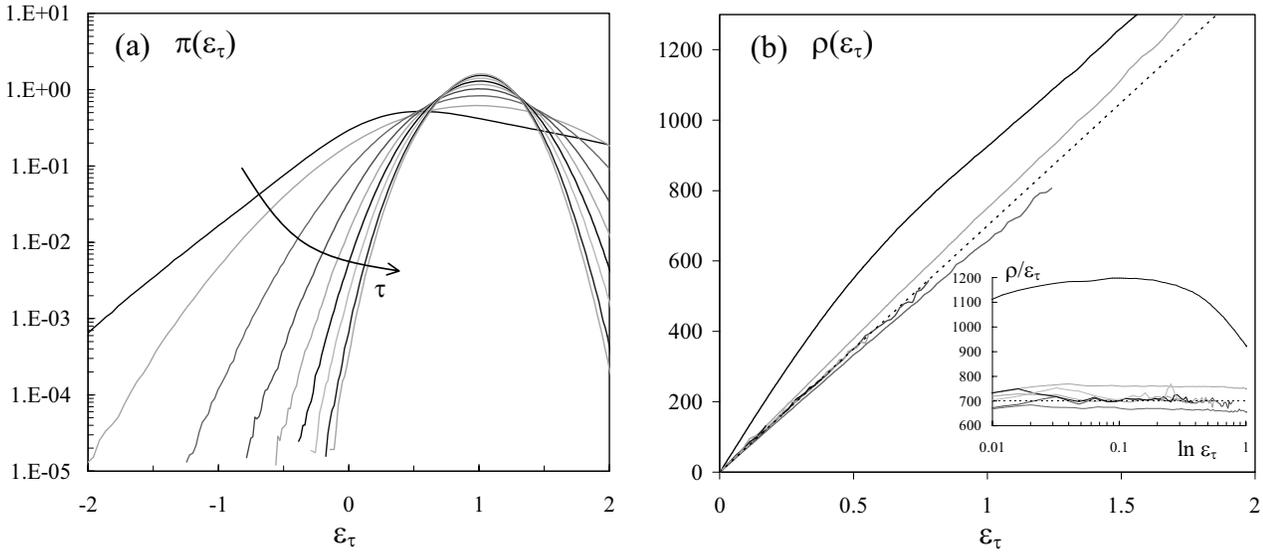}
}
\end{center}
\caption{Periodic forcing.  Statistics of the averaged injected
power $\epsilon_\tau$ on time durations of $\tau$ (in increasing
order as displayed by the arrow 3.5 ms ; 6.5 ; 13.5 ; 20 ; 26.5 ;
33.5 ; 40 ; 47.5 ; 52 ms) . Probability density function
$\pi(\epsilon_\tau)$ (a) and $\rho(\epsilon_\tau)$ (b). Inset:
compensated value $\rho(\epsilon_\tau)/\epsilon_\tau$ in a semi
log plot. In (b) the dashed line corresponds to a linear law of
slope $\delta_P=700$ Hz.} \label{fig:CG-sinus}
\end{figure*}

\begin{figure*}
\begin{center}
\resizebox{2\columnwidth}{!}{%
  \includegraphics{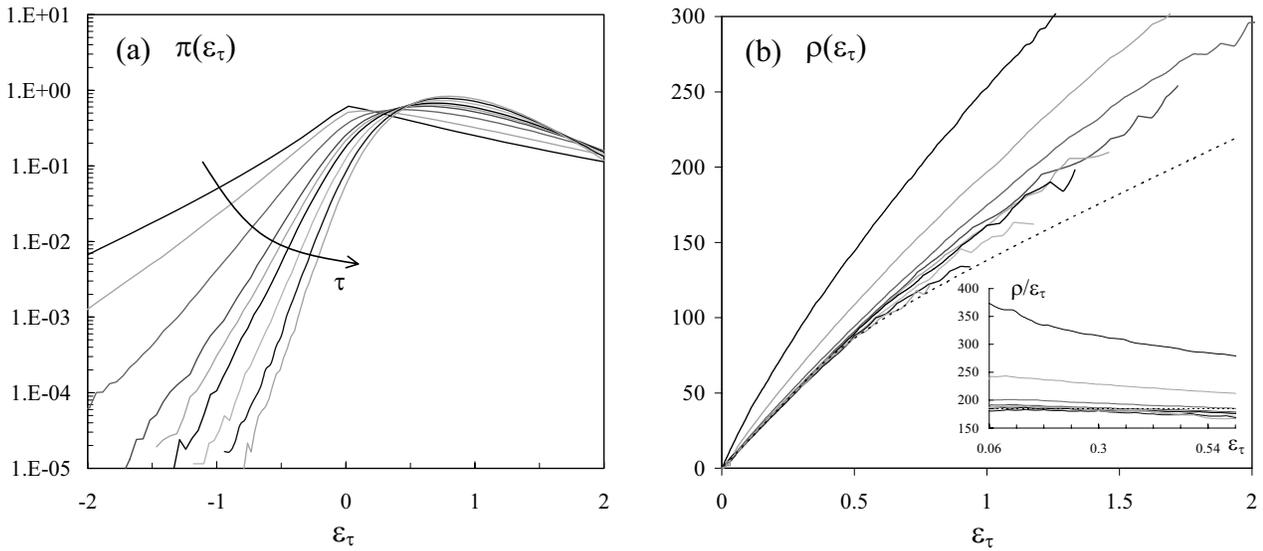}
}
\end{center}
\caption{Random forcing.  Statistics of the averaged injected
power $\epsilon_\tau$ on time durations of $\tau$ (in increasing
order as displayed by the arrow, 6.5 ms ; 13.5 ; 26.5 ; 40 ; 46.5
; 53.5 ; 66.5 ; 80 ; 91.5 ms). Probability density function
$\pi(\epsilon_\tau)$ (a) and $\rho(\epsilon_\tau)$ (b). Inset:
compensated value $\rho(\epsilon_\tau)/\epsilon_\tau$. In (b) the
dashed line corresponds to the theoretical expectation
of~\cite{farago02}: $\rho(\epsilon_\tau)=4\gamma
\epsilon_\tau$ if $\epsilon_\tau<1/3$ or
$\rho(\epsilon_\tau)=7\gamma \epsilon_\tau/4 +3\gamma/2 - \gamma
/(4 \epsilon_\tau)$ if $\epsilon_\tau>1/3$ with $4\gamma = \delta_R
= 184$Hz.} \label{fig:CG-alea}
\end{figure*}

We now consider the (nondimensional) injected work during a time
interval $\tau$,
\begin{equation}\label{epsilon}
    \epsilon _\tau=\frac{1}{\tau}\int_t^{t+\tau}\frac{p(t')}{<p>}dt',
\end{equation}
and the asymmetry function
\begin{equation}\label{SSFT}
    \rho (\epsilon_\tau)=\frac{1}{\tau} \ln
    {\frac{\pi(+\epsilon_\tau)}{\pi(-\epsilon_\tau)}},
\end{equation}
where $\pi(\epsilon_\tau)$ is the probability density function of
$\epsilon _\tau$. The quantity $\rho (\epsilon_\tau)$ is central
for the fluctuation theorem \cite{evans93,gallavotti95,kurchan98},
whose conclusions lead to a linear relationship $\rho
(\epsilon_\tau)=\delta \epsilon_\tau$ for $\tau \rightarrow
\infty$. Here $\delta$ is the contraction rate of the phase space
in the chaotic version of the theorem, or $<p>/\delta$ the
temperature of the stochastic bath in contact with the system. The
pdfs $\pi(\epsilon_\tau)$ are presented for several values of the
integral time duration $\tau$ in Figures~\ref{fig:CG-sinus}(a) and
\ref{fig:CG-alea}(a). For the periodic forcing in
Figure~\ref{fig:CG-sinus}(a), the pdfs converge quickly to a
Gaussian distribution as $\tau$ increases. This is not observed
for the random forcing, for which the pdf remains very different
from the Gaussian even for the largest time duration studied
$\tau=91.5$ ms. In both cases, events of negative power are
observable. We then plot in Figures~\ref{fig:CG-sinus}(b) and
\ref{fig:CG-alea}(b) the quantity $\rho(\epsilon_\tau)$ for each
forcing. For the periodic forcing a linear law
$\rho(\epsilon_\tau)\simeq \delta_P \epsilon_\tau$ is found for
durations $\tau$ larger than 13.5 ms. We can see, in the inset of
Figure~\ref{fig:CG-sinus}(b), that
$\rho(\epsilon_\tau)/\epsilon_\tau$ converges to a unique value
(estimated to $\delta_P = 700 Hz$) which is consistent with the
conclusion of the fluctuation theorem. For the random forcing
presented in Figure~\ref{fig:CG-alea}(b), the relationship between
$\rho(\epsilon_\tau)$ and $\tau$ is not linear. The conclusion of
the fluctuation theorem is not verified. However for large $\tau$,
the curves $\rho(\epsilon_\tau)$ seem to collapse on the nonlinear
asymmetry function predicted in~\cite{farago02}. The asymmetry
curves are approximately linear, $\rho(\epsilon_\tau)\simeq
\delta_R \epsilon_\tau$, only for $\epsilon_\tau<1/3$ as shown in
the inset of Figure~\ref{fig:CG-alea}(b); we find
$\delta_R=184Hz$.
\section{Discussion}
\label{sec:3}

We studied the statistic of the power input in a vibrating plate
set into a chaotic state of wave turbulence by either a periodic
or a random external forcing. The two drivings lead to
fundamentally different distributions for the instantaneous power:
with Gaussian tails in the periodic case and exponential tails in
the random case. A major result is that the conclusion of the
fluctuation theorem is met only in the periodic case, while the
results of~\cite{farago02} hold in the random case. We hereafter
discuss the origin of these differences.

\subsection{Origin of the stochasticity of power injection}
We first try to relate the measurements to the equation of motion
of the plate, as well as to existing models for energy fluxes in
out of equilibrium systems. The vibration can be decomposed onto
the first $N$ eigenmodes of the plate (with $N$ arbitrary large).
Denoting by $x_i(t)$ the modal amplitude, $\Gamma_i$ the modal
damping rate, and $\omega_i$ the associated eigenfrequency, the
dynamics is governed by, for $i=1,...,N$ :
\begin{equation}
\ddot{x_i} + \Gamma_i \dot{x_i} +\omega^2_i x_i= <F,\phi_i> +
<F_T,\phi_i>, \label{pfd}
\end{equation}
where $F(\vec{r}, t)$ is the external force, $F_T(\vec{r}, t)$ is
the internal nonlinear force representing the wave turbulence
feedback, $\phi_i(\vec{r})$ is the eigenmode shape, and $<.\,,.>$
stands for the scalar product associated to the linear operator of
the PDE of motion (von K\'arm\'an model, see e.g.
\cite{touze02,during06}). Because wave turbulence is a chaotic
state, we can reasonably assume that $F_{T}$ is statistically
independent from the forcing $F$, except for the amplitude of
$F_T$, which is prescribed by that of $F$.

Considering identical damping rates for all the excited
eigenmodes, Eq.~(\ref{pfd}) is exactly the starting point of the
stochastic version of the fluctuation theorem~\cite{kurchan98}
where $N$ oscillators, in contact with a thermal bath, are excited
with a deterministic force. Eq.~(\ref{pfd}) should then describe
the experimental case with the periodic forcing, but assuming in
addition that $F_T$ is $\delta$-correlated in time as stated
in~\cite{kurchan98} ; in that context, $F_T$ stands for the effect
of the thermal bath with which the oscillators are in contact.
Here $F_T$ has a finite correlation time which is equivalent to a
$\delta$-correlation in the long time limit, so that the plate
should obey the fluctuation theorem, as we found for
$\epsilon_\tau < 0.75$ (see Figure~\ref{fig:CG-sinus}). The
statistical model presented in section~\ref{sec:Periodic model},
yielding a distribution of instantaneous power with a logarithmic
cusp at zero and Gaussian tails is in good agreement with the
experimental result. This model has been built with the assumption
that the response velocity can be viewed as a sum of a random
noise and a sine function which is consistent with
Eq.~(\ref{pfd}). The random noise term is due to wave turbulence
feedback which dominates the response for the periodic forcing.

Coming back to the full Eq.~(\ref{pfd}), the case such that
$F_{T}=0$ and $F$ is a Gaussian variable was investigated
theoretically in~\cite{farago02,farago04} for one oscillator;
however this Langevin equation has then an unusual meaning as the
random variable plays the role of the forcing and not that of a
thermal bath. Therefore, the conclusions of the fluctuation
theorem are not met~\cite{farago02,farago04} as in our experiment
with the random forcing (Figure~\ref{fig:CG-alea}(b)).
Furthermore, our experimental data seem to approach the
theoretical law of~\cite{farago02,farago04} for the asymmetry
function $\rho$ when $\tau\rightarrow\infty$, for which one obtain
a good collapse for $\epsilon_\tau < 0.75$, as shown in
Figure~\ref{fig:CG-alea}(b). The agreement with the Farago model's
shows that the feedback from wave turbulence might be negligible;
the properties of the injected power could be then simply
interpreted as the result of a linear response of the velocity to
a stochastic force. On the other hand, the reduction of
Eq~\ref{pfd} to a single oscillator might be irrelevant in our
physical context and the extension of Farago's model to $N$
oscillators would be more adequate.
Besides, the velocity response
of the linear dynamical system Eq.~(\ref{pfd}) is also Gaussian,
but correlated to $F$, and the resulting joint pdf of $v$ and $F$
is the bivariate normal distribution with a correlation
coefficient $r=\frac{<vF>}{\sigma_F \sigma_v}=\frac{\sigma_F}{m
\gamma \sigma_v}$~\cite{falcon08,bandi08}, which yields a
distribution of instantaneous power with a logarithmic cusp at
zero and exponential tails. When the plate is randomly forced, we
also find experimentally these distributions (see
Figures~\ref{fig:JointPdf-F,v}(b) and~\ref{fig:Pdf-p}(b)).

\subsection{Phase space contraction}
The relationship between the asymmetry function $\rho$ and the
rescaled work $\epsilon_\tau$ defines a timescale of the power
input fluctuation for each type of forcing. It is surprising to
find the two timescales to be entirely different. In the periodic
case, the slope $\delta_P=700$Hz (Figure~\ref{fig:CG-sinus}b)
should correspond to the phase space contraction, say $\gamma =
700$Hz, while in the random case the theoretical fit in
Figure~\ref{fig:CG-alea}(b) gives a value of the effective damping
rate $\gamma =\delta_R/4=46$Hz.
\begin{figure}
\begin{center}
\resizebox{1\columnwidth}{!}{%
  \includegraphics{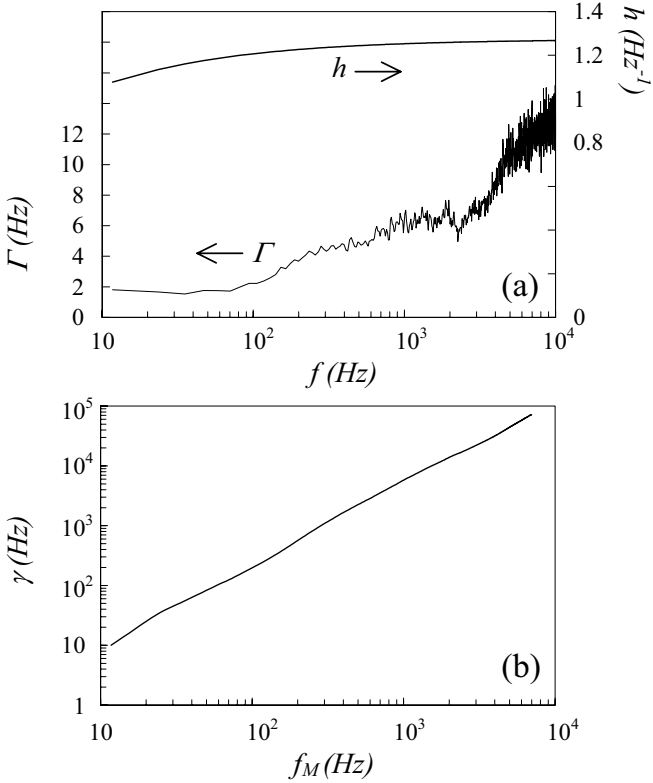}%
}
\end{center}
\caption{(a): modal density $h$ and damping rate $\Gamma$ of the
plate as given in \cite{arcas07}.(b): phase space contraction for
all oscillators of frequencies in the interval $(0,f_M)$ (see
text).} \label{fig:Amortissement}
\end{figure}
In order to understand these values, we may consider the plate as
the set of oscillators corresponding to its excited eigenmodes as
in Eq~\ref{pfd}. Each eigenmode has a well-defined damping rate
$\Gamma_i$ and phase contraction $\gamma$ in the phase space
should be the sum of all damping rates. More quantitatively, let
$h(f)$ be the modal density (the number of eigenmodes of frequency
in the infinitesimal interval $(f,f+\mathrm{d}f)$ divided by
$\mathrm{d}f$) and $\Gamma(f)$ the damping rate for a vibration at
frequency $f$. Then the phase space contraction for all
oscillators of frequencies in the interval $(0,f_M)$ is
\begin{equation}
\gamma(f_M)=\int_0^{f_M} \Gamma(f)h(f)\mathrm{d}f.
\end{equation}
Both $\Gamma(f)$ and $h(f)$ were measured in~\cite{arcas07} for
the same plate and reported here in
Figure~\ref{fig:Amortissement}(a); it turns out that the modal
density is almost constant $h(f)=1.3$ modes/Hz.

Using these measurements, we plot in
Figure~\ref{fig:Amortissement}(b) the phase space contraction
$\gamma(f_M)$ as a function of the maximal frequency $f_M$ of the
oscillators involved. From this figure, the periodic case value
$\gamma(f_M)=700$Hz yields $f_M\approx225$ Hz which roughly
corresponds to the cutoff frequency of wave turbulence (see
Figure~\ref{fig:Spectre-F,v}). This suggests that the effective
system comprises all eigenmodes involved in wave turbulence,
highlighting the chaotic nature of the system. In contrast, the
random case value $\gamma(f_M)=46$Hz yields $f_M\approx31$Hz,
which is half the magnitude of the maximal frequency 75Hz of the
random forcing. This suggests that the effective system is made of
limited number of oscillators. These remarks emphasize the fact
that the nonlinear dynamics is important for the periodic forcing
while the system is dominated by the linear dynamics for the
random forcing.

\subsection{Concluding remarks}
The nature of the forcing turns out to be crucial for the
statistics of power input in a chaotic system. Although we
investigated the two limiting cases of a deterministic, periodic
forcing and a random, Gaussian one, one might expect intermediate
regimes, yet to be explored. We found that the fluctuation theorem
holds in the deterministic case and not in the random one. In the
context of turbulent systems in general, it is not clear how
useful would be the fluctuation theorem. Actually, the energy flux
from the large scales to the dissipative scales might be dominated
by the randomness of the force applied by the large scales. To
clarify this issue, it is necessary to determine how much
randomness of the forcing is necessary to depart from the
fluctuation theorem. To do so, much more experimental and
theoretical work is needed. On the one hand, it would be
interesting to compare the two types of forcing in other turbulent
systems. On the other hand, the random forcing of a thermostated
system does not seem to have received theoretical attention so
far.

\subsection*{Acknowledgements}
The authors are grateful to K. Arcas for fruitful discussions and
for providing his measurements presented in
Figure~\ref{fig:Amortissement}.

\end{document}